\documentclass[amsmath,amssymb, aps, prl, twocolumn]{revtex4-1}

\usepackage{dcolumn}
\usepackage{array}
\usepackage{bm}
\usepackage{amsmath}
\usepackage{amssymb}
\usepackage{graphicx}
\usepackage{esint}
\usepackage{txfonts}
\usepackage{wasysym}
\usepackage{subfigure}
\usepackage{color}
\usepackage{array}
\usepackage{epsfig}

\begin{document}

\title{Machine Learning $\mathbb{Z}_{2}$ Quantum Spin Liquids with Quasi-particle Statistics}

\author{Yi Zhang$^1$}
\email{frankzhangyi@gmail.com}
\author{Roger G. Melko$^{2,3}$}
\author{Eun-Ah Kim$^1$}
\email{eun-ah.kim@cornell.edu}
\affiliation{$^1$ Department of Physics, Cornell University, Ithaca, New
York 14853, USA}
\affiliation{$^2$ Perimeter Institute for Theoretical Physics, Waterloo, Ontario N2L 2Y5, Canada}
\affiliation{$^3$ Department of Physics and Astronomy, University of Waterloo, Ontario, N2L 3G1, Canada}

\date{\today}

\begin{abstract}
After decades of progress and effort, obtaining a phase diagram for a strongly-correlated topological system still remains a challenge. Although in principle one could turn to Wilson loops and long-range entanglement, evaluating these non-local observables at many points in phase space can be prohibitively costly. With growing excitement over topological quantum computation comes the need for an efficient approach for obtaining topological phase diagrams. Here we turn to machine learning using quantum loop topography (QLT), a notion we have recently introduced. Specifically, we propose a construction of QLT that is sensitive to quasi-particle statistics. We then use mutual statistics between the spinons and visons to detect a $\mathbb Z_2$ quantum spin liquid in a multi-parameter phase space. We successfully obtain the quantum phase boundary between the topological and trivial phases using a simple feed forward neural network. Furthermore we demonstrate advantages of our approach for the evaluation of phase diagrams relating to speed and storage. Such statistics-based machine learning of topological phases opens new efficient routes to studying topological phase diagrams in strongly correlated systems.
\end{abstract}
\maketitle

\section{Introduction}
Despite much interest in topological phases of matter, the search for and detection of the finite regions of phase space that support topological order has been a long standing challenge. This is a nontrivial challenge because microscopic models of strongly-correlated topological order are usually established in exactly solvable models at first \cite{Kitaev20032, Kitaev20062, Haah2011, ChenSPT, Wen2016, Chamon2005,Hong2007}. Nevertheless, universal properties of topological phases from low-energy effective theories, i.e. topological quantum field theories, provide a broader support for the understanding of
topological phases. Naturally, much effort has gone into perturbing exactly solvable models both theoretically  \cite{fracton2017,fracton2016, Karimipour2013, Fradkin1979, Vidal2009, Razieh2015} and numerically \cite{Vidal2011,Stamp2010,Trebst2007, GENOVESE2003,Jongeward1980,Claudio2008, Vidal2012} to investigate the stability of the associated topological phases beyond the fine-tuned solvable points. Moreover, growing enthusiasm over the idea of using exotic statistics of excitations for topological quantum computation \cite{Kitaev20032, Chetan2008, Haah2011} and the experimental quests driven by related proposals \cite{Roman2010,Oreg2010} have raised the need to understand the stability of various topological phases and establish the corresponding phase diagrams.

It is perhaps the key interesting feature of topological phases that simultaneously underlies challenges in their numerical diagnosis: the absence of local order parameter. Nevertheless various measures of non-local
correlations have enabled progress in evaluating phase diagrams and detecting phase transitions. Among the most successful approaches are expectation value of Wilson loop\cite{Scalapino2002,Wen2008tqpt}, and entanglement entropy\cite{Claudio2008, Pollmann2014, KitaevTee, WenTee}. Unfortunately, the non-local nature of the respective estimators can make the algorithms for measuring these costly in general. In addition, in some cases one can use a thermodynamic signature such as specific heat to detect the phase transition \cite{Pollmann2014, Subir2002, Stamp2010, Trebst2007}. Although a singularity in specific heat is an effective indicator of a phase transition, it has the drawback that it does not reveal any information regarding the topological aspects of the associated phases. Hence, in addition to these standard techniques, developing a cost-effective approach that can map out a phase diagram with topological quantum phase transitions using key features of the topological phase, such as non-trivial statistics, is highly desirable.

A new strategy for a dramatic speed-up in the approximate evaluation of phase diagrams is to use neural-network based machine learning \cite{MLbook}.
Efforts in this direction fall into one of two broad categories: unsupervised learning and supervised learning.
In the unsupervised context, the task of classifying raw state configurations with phase labels (e.g.~``clustering'' to find hidden patterns or grouping in data) is one actively-pursued goal.  Several different approaches have been used, including principal component analysis and neural networks \cite{LeiWang2016,Nieuwenburg2017,Hu2017}, resulting
in a rapidly-developing sub-field.
Within the supervised learning approach, the algorithmic strategy is perhaps more well established. There, neural networks can be trained with data in the form of raw state configurations, each labelled by its respective phase. Once the neural network is trained, a new (``test'') data set is given to it, and it is tasked with labelling each configuration with one of the phases it has been trained to recognize. This approach has been successful in obtaining phase diagrams with conventional ordered phases\cite{Melko20161, Simon2016, Kelvin2016}, many-body localizations\cite{Titus2017}, or chiral topological phases\cite{Ohtsuki2016,qlt2016,Ohtsuki2017}. Nevertheless, identifying non-chiral topological order remains surprisingly challenging for such supervised machine learning approaches\cite{Melko20161}.

Here we propose a learning strategy based on the non-trivial statistics between fundamental excitations -- the key defining property of correlated topological phases. This approach is inspired by earlier efforts to calculate quasi-particle statistics that rely on evaluating quantities reflecting long-range entanglement of correlated topological phases \cite{smat,smat2,Frank2013prb, Frank2014prb,Meiwen2015,Zaletel2013,Vidal2013,Melko2015}. Since such calculations are typically computationally costly,
attempting to sweep a large phase space can become prohibitively difficult.
Hence we propose using a technique called quantum loop topography (QLT) \cite{qlt2016}, designed around quasi-particle statistics in conjunction with neural-network based machine learning. The notion of the QLT was first introduced in Ref.~\cite{qlt2016} as a preprocessing interface between a quantum many-body state and a feed forward neural network. The key idea was to select and organize the input data based on the physical response defining the target phases to detect. This method was successfully implemented for chiral topological phases in Ref.~\cite{qlt2016}. Here we build a QLT protocol for non-chiral topological phases with non-trivial quasi-particle statistics. We then demonstrate the effectiveness of our QLT-based machine learning strategy by obtaining the phase diagram of the toric code in the presence of magnetic fields, mapping out the $\mathbb Z_2$ quantum spin liquid in the parameter space of the external field strengths.

The rest of this paper is organized as follows. In section II, we present the general strategy for evaluating phase diagrams involving topological phases using neural network based supervised machine learning with quantum loop topography. In Sec. III, we discuss the model of field-perturbed toric code and the QLT guided by the quasi-particle mutual statistics and specified for the detection of the $\mathbb Z_2$ quantum spin liquid phase. In Sec. IV, we present the architecture of our neural network and the algorithm for performing supervised machine learning. In Sec. V, we present the results bench-marked against conventional specific heat collapse method. We then close with a discussion and outlook.

\section{Quantum loop topography for machine learning}\label{sec:mlqlt}

\begin{figure}
\includegraphics[scale=0.35]{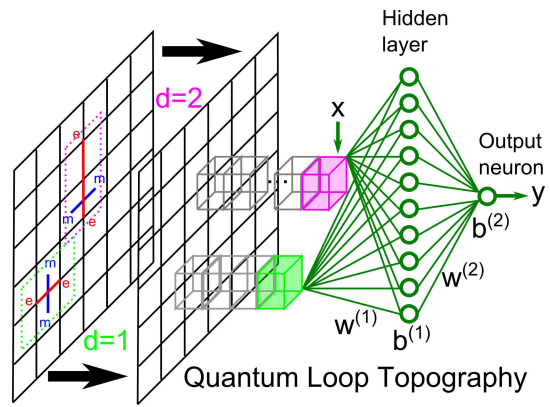}
\caption{Schematic illustration of our machine learning architecture. Quantum loop topography uses an ensemble of `minimally non-local' operators to extract information from the many-body systems and creates `image' with relevant information for the input into the artificial neural network.}
\label{fig:machine}
\end{figure}

The evaluation of a phase diagram using neural network based supervised machine learning is a process made of three key stages \cite{Melko20161}. The first stage is to assemble diverse groups of input data representing each of the phases of interest. This collection of data sets forms the ``training set''. Each element of the training set is labeled with the respective phase. The second stage is to construct the neural network and train it to categorize the training set correctly by providing feedback with the known label. During the training process the neural network adjusts the parameters that define the network (the weight matrix $w$ and bias vector $b$) to minimize the error in its output $y$ (see Fig.~\ref{fig:machine}). The final stage is to sweep through the phase space and a ``test'' data set for each phase space point is analyzed by the neural network. Such efficient and automated study of the entire phase space is one important merit of the neural-network based approach and is fueling rapidly increasing efforts in this direction\cite{Melko20161,Simon2016, Kelvin2016,Ohtsuki2016,qlt2016}.

Quantum loop topography was introduced in Ref.~\cite{qlt2016} after it became clear through previous efforts \cite{Simon2016} that for efficiency or even success in training, it is critically advantageous to use input data containing information relevant to the target phases. QLT is a data preparation (or preprocessing) stage that builds on the characteristics of a phase \cite{qlt2016}. The necessity for QLT arises when local information is insufficient for the phase identification as in topological phases (and in superconducting phases). Traditionally, one would either use maximally non-local information such as entanglement entropy or explicitly evaluate the relevant response function. Using QLT combined with machine learning, one avoids both of these time consuming approaches while keeping their key elements. A QLT is a ``topographic'' map made of un-averaged and therefore fluctuating values of products of loop-forming operators that are relevant for the phase of interest. By retaining only the loops of sizes below certain cutoff, i.e. minimally non-local loops,
a QLT can readily take advantage of the short correlation lengths in gapped phases, nevertheless incorporating non-local correlations.

The key to success for any QLT is to involve operators relevant for the defining properties of the target phase and this approach is particularly useful when usual equilibrium order parameter is not computationally accessible.
Earlier, two of us introduced the notion of QLT and successfully implemented it targeting integer and fractional Chern insulator~\cite{qlt2016}. For such chiral topological phases, the natural response function to guide the construction of QLT was the Hall conductivity. On the other hand, of increasing importance are non-chiral topological phases with multiple types of fractional excitations.
Here we focus on such non-chiral topological phases when one cannot employ the Chern number inspired QLT from Ref.~\cite{qlt2016}. The new defining propoerty we will focus on is the quasi-particle statistics.



One of the defining properties of intrinsic topological phases is the non-trivial exchange and braiding statistics that their quasi-particle excitations must obey. Such statistical information is encoded into the topological quantum field theory associated with a given topological phase through the expectation values of Wilson loops forming non-trivial knots. When the world-line of a quasi-particle forms a knot with that of another quasi-particle, such a knot can only be resolved through an appropriate unitary transformation in the quasi-particle Hilbert space \cite{Witten1989,Kitaev20032,Nayak2008}. To be specific, let us consider a gapped Abelian topological phase. The long-distance (IR) effective theories associated with such phases are  Chern Simons
theories with $K$-matrices \cite{Wen1992,Read1990}:
\begin{equation}
\mathcal{L}_{CS}=\frac{K_{IJ}}{4\pi} \epsilon^{\mu\nu\lambda} a^I_\mu
 \partial_{\nu} a^J_{\lambda} - a^I_\mu j^\mu_I
\label{eq:kmat}
\end{equation}
where $a^I_\mu$ are $U(1)$ gauge fields coupled to quasi-particle
currents $j_I^\mu$, $I=1,2,\cdots$ labels the fundamental types of
quasi-particle excitations. For instance,
the $K$ matrix associated with Laughlin states at filling $\nu=1/m$ has a single entry $K_{11}=m$ while that associated with $\mathbb{Z}_2$ quantum spin liquid is
\begin{equation}
K=\left(\begin{array}{cc}
0 & 2\\
2 & 0
\end{array}\right).
\end{equation}

\begin{figure}
\includegraphics[scale=0.25]{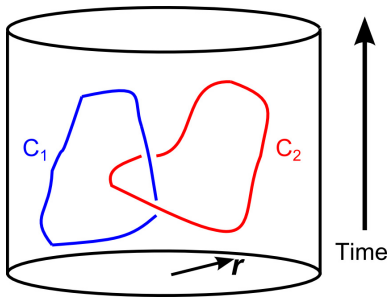}
\caption{A nontrivial knot with linking number $L=1$ between two Wilson loops $C_1$ and $C_2$ in 2+1-dimension space-time.}
\label{fig:loopknot}
\end{figure}

From the field theory perspective, the Wilson loops formed by the quasi-particle (quasi-hole) trajectory $C$ defined by
\begin{equation}
W^I_{C}\equiv\mathcal{P}\exp\left(i\oint_{C} a^I_l
dl\right),
\end{equation}
where $\mathcal{P}$ is the path-ordering symbol,
are the only gauge-invariant observables.
The quasi-particle statistics are encoded in Wilson loops through \cite{Witten1989,Kitaev20032,Nayak2008}:
\begin{equation*}
\mathcal{P} W^I_{C_1} W^J_{C_2} = e^{2\pi i K^{-1}_{IJ} L\left(C_1, C_2\right)}
\end{equation*}
where $L\left(C_1, C_2\right)$ is the space-time linking number between the $C_1$ and $C_2$ loops.
For example, we illustrate a two-Wilson-loop configuration forming a nontrivial knot ($L=1$) in Fig. \ref{fig:loopknot}.
In a topological quantum field theory, the expectation values of Wilson loop configurations only depend on the topology irrespective of the size of Wilson loops involved, since a topological quantum field theory is essentially scale-free.

Nevertheless when topological phases are investigated through lattice models, microscopic details and short-distance (UV) physics enter the problem and long-distance (IR) physics will  only appear at length scales much longer than the short-distance cutoff. As a result, previous numerical studies focused on the largest loops the system can support, e.g. the two cycles of a torus \cite{smat,smat2,Zaletel2013,Vidal2013}.
Unfortunately, trying to numerically evaluate such global objects with precision is highly costly. Clearly, the statistical information would be present in the knots of smaller size as well \cite{Witten1989,Nayak2008,Wen2005prb}, such as the one in Fig.~\ref{fig:lattice}b. However, the short-distance details and perturbations prevent us from reaching comparative theoretical predictions and making use of the information from such smaller knots, even though they can be much easier to study numerically.

Instead, we now propose using QLT based upon these semi-local knots to select and build the input data, which is analyzed from a novel perspective using machine learning with an artificial neural network (see Fig. \ref{fig:machine}). When supervised machine learning is successfully accomplished, the resulting architecture can scan the phase space and obtain the phase regions of the topological phases efficiently and automatically.

\section{$\mathbb Z_2$ quantum spin liquid and quantum loop topography}\label{sec:z2qsl}

The $\mathbb Z_2$ quantum spin liquid is the prototypical example of a state
with a non-chiral topological order defined on a lattice \cite{Kitaev20032, White2010} . It is a strongly-correlated quantum spin liquid, with four-fold ground-state degeneracy separated from the excited states with a full gap when defined on a torus. Its non-chiral nature and the lack of topological edge states make the detection of $\mathbb Z_2$ quantum spin liquid even more elusive. Importantly, its fundamental types of quasi-particle excitations, the spinon and the vison, both have trivial self statistics yet semionic mutual statistics: the system picks up an overall phase factor $e^{i\theta}$ with statistical angle $\theta=\pi$ upon braiding a spinon around a vison, or vice versa.

\begin{figure}
\includegraphics[scale=0.25]{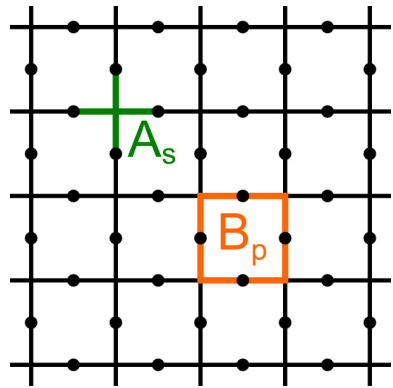}
\caption{Illustration of the lattice spin model in Eq. \ref{eq:ham2d}. The spin-$1/2$'s reside on the bonds of the two-dimensional square lattice. $A_s$ and $B_p$ are the products of $\sigma^x$ and $\sigma^z$ operators around a site $s$ and on the boundaries of a plaquette $p$, respectively.} \label{fig:toric}
\end{figure}

A model known to support a $\mathbb{Z}_2$ quantum spin liquid phase is \cite{Stamp2010, Vidal2009, Vidal2012}:
\begin{equation}
H_{2D}=-J_x\sum_s A_s - J_z \sum_p B_p -  h_x \sum_j \sigma_j^x - h_z \sum_j  \sigma_j^z
\label{eq:ham2d}
\end{equation}
where the spin-$1/2$ lives on the bonds of a square lattice and  $A_s=\prod_{j\in s} \sigma^x_j$ and $B_p=\prod_{j\in p} \sigma^z_j$ are the products of spin operators around a site $s$ and on the boundaries of a plaquette $p$, respectively [see Fig. \ref{fig:toric}]. The $h_z$ and $h_x$ terms are external magnetic fields in the $\hat x$ and $\hat z$ directions, respectively. For the rest of the paper, we consider system size $12\times12$ unless noted otherwise.

\begin{figure}
\includegraphics[scale=0.40]{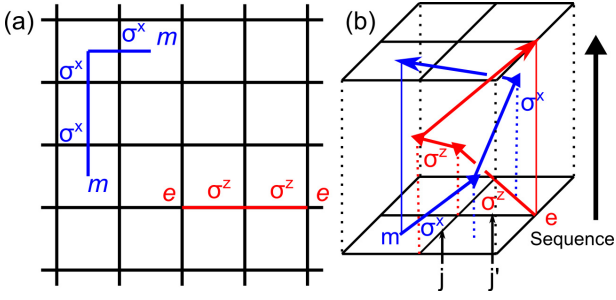}
\caption{(a) A string of $\sigma^z$ ($\sigma^x$) operators creates at its two ends a pair of spinons (visons) denoted as `e' (`m'), or equivalently, moves one spinon (vison) from one end to the other. Note that the spinon string $C_e$ and vison string  $\tilde C_m$ live on the bond of the original and dual square lattices, respectively. (b) Illustration of a nontrivial knot between the world-lines of a spinon and a vison as a discrete lattice version of Fig. \ref{fig:loopknot}.} \label{fig:lattice}
\end{figure}

At the special point $h_x=h_z=0$, $H_{2D}$ amounts to Kitaev's toric code \cite{Kitaev20032}. The exact solvability of the toric code has allowed for much progress in explicit understanding of $\mathbb Z_2$ quantum spin liquid. Since all $A_s$ and $B_p$ commute in the Hamiltonian, the ground states are simply given by allowing $A_s=B_p=1$, while the quasi-particle excitations are associated with violations that cost finite energy penalties: a spinon at site $s$ with $A_s=-1$ and a vison in plaquette $p$ with $B_p=-1$. Note that a string of $\sigma^z$ ($\sigma^x$) operators $\prod_{j\in C} \sigma^z_j$ ($\prod_{j\in \tilde C} \sigma^x_j$) moves one spinon (vison) from one end of $C$ ($\tilde C$) to the other end, see Fig. \ref{fig:lattice}a. To illustrate the mutual statistics, consider the process $\prod_{j\in p} \sigma^z_j$ that circles a spinon around a plaquette $p$: it gives rise to a phase factor $B_p$, which is $-1$ if there exists a vison inside the plaquette.

The toric code is, however, non-generic - its correlation length is zero and quasi-particles are strictly not allowed in the ground states. We consider presence of magnetic fields: the $h_z$ ($h_x$) magnetic field makes it more preferential to generate spinon (vison) pairs, which will eventually condense and give rise to a spin-polarized phase beyond a critical field strength. The two topologically-trivial phases, usually called magnetically ordered phase and disordered phase at either large $h_z$ or large $h_x$ are dual to each other under $\sigma_x \leftrightarrow \sigma_z$ transformation.

In practice, Eq. \ref{eq:ham2d} can be mapped to a three-dimensional classical system - the (anisotropic) $\mathbb Z_2$ gauge Higgs model \cite{Fradkin1979} via imaginary time evolution and Trotter decomposition \cite{Stamp2010}. After separating the imaginary time into a large number of small intervals $\beta=n\Delta \tau$ the operator $\exp\left(-\beta H\right)$ for the quantum partition function can be approximated by $\left[\exp\left(-\Delta\tau H_x\right)\exp\left(-\Delta\tau H_z\right)\right]^n$. $\beta=1/k_B T$ is the inverse temperature, and $H_z$ ($H_x$) are the terms with $\sigma_z$ operators ($\sigma_x$ operators). This maps the partition function to a three-dimensional system with $n$ lattice spacings along the imaginary time direction, and Ising spin-$1/2$ denoted as $S_j$ lives on the cubic lattice bonds. The statistical weight of a given $\{S_j\}$ configuration on each imaginary time slice and between adjacent slices is determined by $H_z$ and $H_x$, respectively. After careful treatment of the gauge redundancy, the  $J_x$, $h_z$, $h_x$ and $J_z$ terms in Eq.~\ref{eq:ham2d} are interpreted as the vertical bond, horizontal bond, vertical plaquette and horizontal plaquette terms in the anisotropic $\mathbb Z_2$ gauge Higgs model. For simplicity, we consider the isotropic case where
\begin{equation}
\beta  H_{3D}=-\lambda_b\sum_j S_j -\lambda_p\sum_p\prod_{j\in p}S_j
\label{eq:ham3d}
\end{equation}
where dimensionless parameters $\lambda_b$ and $\lambda_p$ are related to the parameters ot the model Eq.~\eqref{eq:ham2d} through
\begin{eqnarray}
\lambda_b = h_z \Delta\tau = -\frac{1}{2}\ln \tanh J_x \Delta\tau \nonumber\\
\lambda_p = J_z \Delta\tau = -\frac{1}{2}\ln \tanh h_x \Delta\tau
\label{eq:paramap}
\end{eqnarray}
in the limit of small $\Delta \tau$. We note that the method we will present also offers an alternative and convenient signature of de-confinement \cite{Sondhi2011njp} in an equivalent lattice gauge theory, even though our focus here is mainly on the $\mathbb{Z}_2$ quantum spin liquids. For the rest of the paper, we use $\lambda_b$ and $\lambda_p$ as parameters, sticking to the notation of Ref.~\onlinecite{Stamp2010} for benchmark purposes.

The three-dimensional classical model in Eq.~\ref{eq:ham3d} offers a convenient way to measure operators within QLT, which are sampled within classical Monte Carlo Metropolis with both on-site and cluster updates:
\begin{eqnarray}
\left\langle\hat O\right\rangle ={\mbox{tr}\left[\rho_{2D}\hat O\right]}=\sum_{\alpha\gamma}\rho^\gamma_{\alpha\alpha} \left\langle\hat O\right\rangle_\alpha
\end{eqnarray}
where $\rho_{2D}= \sum_{\alpha\beta\gamma} \rho^\gamma_{\alpha\beta} \left|\alpha\right\rangle\left\langle\beta\right|$ is the quantum density matrix of the original two-dimensional system in Eq. \ref{eq:ham2d}. $\alpha$ and $\beta$ are two-dimensional spin-$1/2$ configurations. $\rho^\gamma_{\alpha\beta}=\exp\left(-E_{\alpha\beta\gamma}\right)/Z$ is the thermal statistical weight of the three-dimensional classical configuration with open boundary conditions $\alpha$ and $\beta$ along the imaginary time direction and bulk configuration $\gamma$, whose energy $E_{\alpha\beta\gamma}$ is given by Eq. \ref{eq:ham3d}. Similarly, $\rho^\gamma_{\alpha\alpha}$ is the normalized and positive-definite weight with periodic boundary condition $\alpha$, serving as the sample probability in our Monte Carlo metropolis. Then the quantity that contributes to the operator expectation value is
\begin{equation*}
\left\langle\hat O\right\rangle_\alpha = \sum_\beta \left\langle\alpha\left|\hat O\right|\beta\right\rangle \cdot \rho^\gamma_{\alpha\beta}/ \rho^\gamma_{\alpha\alpha}=\sum_\beta \left\langle\alpha\left|\hat O\right|\beta\right\rangle \cdot \exp\left(-\Delta  E_{\alpha\rightarrow\beta}\right)
\end{equation*}
where $E_{\alpha\rightarrow\beta}$ is the energy difference.

Since the $\mathbb Z_2$ topological order has trivial quasi-particle self statistics, we focus on the mutual statistics and consider for QLT only operators with nontrivial knots between the spinon and vison world-lines. Namely, these are strings of $\sigma^z$ and $\sigma^x$ operators with double intersections. The example illustrated in Fig. \ref{fig:lattice}b corresponds to the operator $\left\langle \sigma_j^x \sigma_{j'}^z \sigma_{j'}^x \sigma_j^z \right\rangle= \mbox{tr} \left[\rho \sigma_j^x\sigma_j^z \sigma_{j'}^z \sigma_{j'}^x  \right]$, where the vison and spinon trajectories intersect twice at $r$ and $r'$, before the density matrix $\rho$ weighs the quasi-particle correlations between the initial and final positions and effectively closes the trajectories.

In the spirit of machine learning, instead of calculating their full expectation values, which are relatively expensive due to averaging over the Markov chain, we settle with an individual Monte Carlo sample $\alpha$ for QLT. To further simplify the QLT, we consider instead the operators $\prod_{j\in C_e} \sigma^z_j \prod_{k\in \tilde C_m} \sigma^x_k$ where the strings $C_e$ and $\tilde C_m$ intersect. When valued at a particular Monte Carlo step, the original operators within QLT can be straightforwardly derived from the ensemble of such intersecting $\prod_{j\in C_e} \sigma^z_j \prod_{k\in \tilde C_m} \sigma^x_k$, making the latter an equally informative candidate for machine learning with QLT. Finally, string tension and the rapidly decaying spinon-spinon and vison-vison correlations allow us to focus on strings no longer than a cut-off, which we choose to set to $d_c=2$ unless noted otherwise.

\section{Machine Learning Topological Phase diagram}\label{sec:algorithm}

Now we describe the specific procedure for studying the phase diagram of the model in Eq.~\eqref{eq:ham3d}. Since the large $\lambda_p$ and small $\lambda_b$ limit and the small $\lambda_p$ and $\lambda_b$ limit are well established for the $\mathbb Z_2$ topological phase and the trivial phase, respectively, they offer the labeled samples needed for supervised training with target neural output $y=1$ and $y=0$, respectively. We illustrate two parallel machine learning approaches with merits of their own. In the efficiency-focused approach, we train a {\it universal} neural network with a diverse training set of multiple points within the known phase space, and then apply the single trained network to the entire $(\lambda_p, \lambda_b)$ phase space. In the precision-focused approach,
we train a separate neural network for each $\lambda_b$-slice with two sample points at large and small $\lambda_p$, and then apply the network to the intermediate phase space on the given $\lambda_b$-slice.

For each value of $\lambda_p$ and $\lambda_b$ chosen for training, we construct 10000 QLT data sets sampled through a classical Monte Carlo Metropolis procedure. We also reserve a separate test set of $20\%$ of the size of the training set for validation purposes including learning speed control and termination \cite{MLbook}. As illustrated in Fig.~\ref{fig:machine}, once obtained, each data set serves as an ``image'' for a fully-connected feed-forward neural network with one hidden layer consisting of $n=20$ sigmoid neurons. Each neuron processes the input through independent weights and biases $w\cdot x+b$. After the hidden layer, the outcome is fed forward to be processed by the output neuron. The final output $0 \le y \le 1$ corresponds to the neural network's judgment of whether the QLT input state is topological. The neural network is trained via back-propagation and gradient descent to optimize the weights and biases. We use the cross entropy as the cost function, with L2 regularization to avoid over-fitting, and a mini-batch size of 10 \cite{MLbook}.

Once the neural network is trained successfully, it can rapidly process QLT data obtained from different parts of the phase space to yield a phase diagram. In order to establish level of confidence on the trained network's assessment of whether the system is topological or not, we process 2000 inputs at each point to obtain statistics. Specifically, for each phase space point $(\lambda_p,\lambda_b)$, we find the fraction $p(\lambda_p,\lambda_b)$ of test inputs with `topological' output, i.e., $y>0.5$. To further suppress the uncertainty due to sample fluctuations, we base our judgment upon the average neural output $\bar y$ over 5 QLT inputs from uncorrelated Monte Carlo samples: a `topological' output corresponds to $\bar y>0.5$; otherwise, the $\bar y<0.5$ output is considered `trivial'. See Supplemental Materials for details.

The extreme corners of the phase diagram in the space of $(\lambda_p,\lambda_b)$  have been schematically understood for a long time\cite{Fradkin1979}. First of all, the topological phase is expected in the region with large $\lambda_p$ and small $\lambda_b$, with the $\lambda_p\rightarrow \infty$ and $\lambda_b \rightarrow 0$ limit equivalent to the zero-field toric code. Moreover, magnetically ordered phase is expected in the large $\lambda_b$ limit, and the disordered phase is expected at the high temperature limit of small $\lambda_p$ and $\lambda_b$. The two topological trivial phases are dual to each other with a self-dual line $\lambda_b = -0.5 \ln\tanh \lambda_p$ as their phase boundary. Numerical pursuits of the phase diagram with increased system sizes and computational efforts\cite{Jongeward1980,GENOVESE2003,Trebst2007,Stamp2010}
established the phase boundary between a disordered (magnetically ordered) phase and topological phase at around $\lambda_p \alt 0.76$ ($\lambda_b \agt 0.223$) with the critical $\lambda_p$ value from the topological phase toward the disordered phase displaying a very slight negative dependence upon increasing $\lambda_b$. The goal of our QLT-based machine learning approach is to reproduce these known results in a way that can be extended to other models whose phase diagrams remain a open question in the future.

\begin{figure}
\includegraphics[scale=0.40]{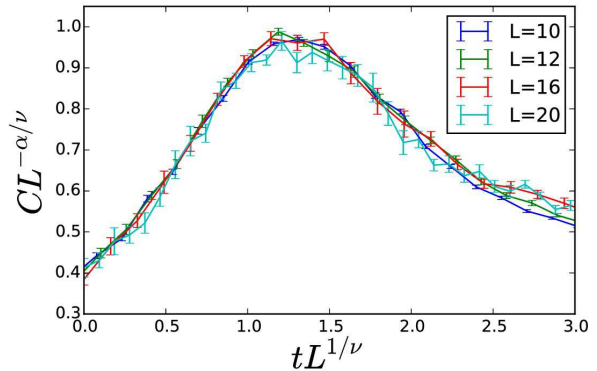}
\caption{Specific heat data of the three-dimensional system $H_{3D}$ from Eq. \ref{eq:ham3d} for a range of system sizes $L=10,12,16,20$ is collapsed according to the finite-size scaling using critical exponents of the three-dimensional Ising universality class. The best collapse is achieved at critical temperature $k_B T_c=1.324(1)$ for $\lambda_p/\beta = 1$ and $\lambda_b/\beta=0.29$.}\label{fig:benchmark}
\end{figure}

In order to precisely benchmark the outcome, we also determine one point of the phase boundary
with a traditional techniques of collapsing specific heat data\cite{Fisher1972}.
This approach requires calculating specific heat at over a mesh of phase space points repeatedly for a range of system sizes and then finding the phase space point that offers the best critical scaling collapse. Specifically, in order to locate the critical point intersected by the line $\lambda_b/\lambda_p=0.29$ in the $(\lambda_p,\lambda_b)$ phase diagram,  we obtain the specific heat data of the classical system in Eq.~\ref{eq:ham3d} with $\lambda_p/\beta=1$ and $\lambda_b/\beta=0.29$ and a variable temperature $k_B T=\beta^{-1}$ for system sizes $L=10,12,16,20$ via conventional Monte Carlo calculations.
We then attempt to collapse the data to the finite-size scaling function using critical exponents of the expected three-dimensional Ising universality class for each of different temperatures, looking for the best data collapse.
Since we find the best data collapse
at temperature $k_B T=1.324$ (see Fig. \ref{fig:benchmark})
with a noticeable deviation nearby  temperatures
as close as $k_B T=1.323$ and $k_B T=1.325$, we conclude $k_B T_c=1.324\pm 0.001$. This locates one critical point in the phase diagram at $(\lambda_p,\lambda_b)=(0.7553\pm0.006,0.2192)$, see Fig. \ref{fig:pd1ds}.  Accuracy of this magnitude comes with a cost, and the Monte Carlo computation involved in producing the large number of samples needed to reduce statistical errors involved about two core-years of CPU time in this example. Unfortunately, this level of accuracy is necessary when the varying quality of scaling collapse at different points in the phase space is used to single out the critical point. Moreover, a priori knowledge of the critical exponents is required in this approach.
Although viable in principle, mapping out the entire two-parameter phase diagram to this accuracy, where this Monte Carlo process needs to be repeated over many other cuts, would quickly incur insurmountable costs\footnote{Note the use of Binder cumulant, which is often preferred over specific heat collapse for finding critical points for its efficiency, requires an order parameter and hence it is not an option for any topological order.}.

\section{Results}\label{sec:results}

\begin{figure}
\includegraphics[scale=0.32]{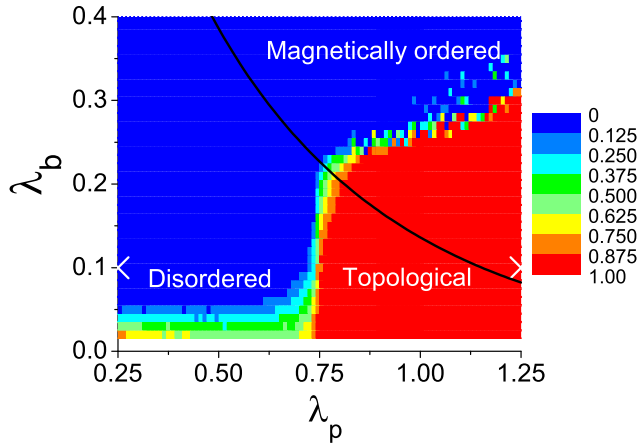}
\caption{Phase diagram in both $\lambda_b$ and $\lambda_p$ obtained with a single, universal neural network with supervised machine learning and quantum loop topography. The color scale indicates the fraction $p(\lambda_p, \lambda_b)$ of topological response: red (blue) color is the topological (trivial) phase. The white crosses denote the training sets. The solid black line labels the self-dual parameters $\lambda_b=-0.5\ln\tanh \lambda_p$.} \label{fig:pd2d}
\end{figure}

First we attempt to obtain the full two-dimensional phase diagram with a single {\em universal} neural network trained with data sets from only two phase-space points: $(\lambda_p,\lambda_b)$ = (0.25,0.1) for the non-topological phase and (1.25,0.1) for the topological phase. The resulting phase diagram shown in Fig.~\ref{fig:pd2d} reproduces the known phase diagram reasonably well, especially taking into account the rapid speed of the neural network scanning for the $\mathbb{Z}_2$ topological phase. Specifically, the entire process of training set preparation, neural network machine learning, and eventually scanning a phase space of $101\times 39$ pairs of $(\lambda_p, \lambda_b)$ parameters takes just about 4 hours of CPU time. The lack of diversity in the training set as a result of including data from only two points in phase space \footnote{One way to improve the neural network's performance is to increase the diversity of the training set and include more phase space.} affects the reliability of the neural network for small $\lambda_b < 0.05$, and also near the tri-critical point around $(\lambda_p, \lambda_b)\sim(0.75,0.225)$. Nevertheless one remarkable outcome of this result (obtained from a single neural network trained with only one disordered phase-space point and one topological phase-space point) is the apparent understanding of the duality. Specifically, although the network was trained deep in the disordered phase as the only topologically trivial example, it correctly recognized the magnetically ordered phase also as a topologically trivial phase. Indeed, this suggests that the decision-vital information is not specific to spinon or vison alone, but between spinon and vison and invariant under the duality transformation, such as their mutual statistics.

\begin{figure}
\includegraphics[scale=0.32]{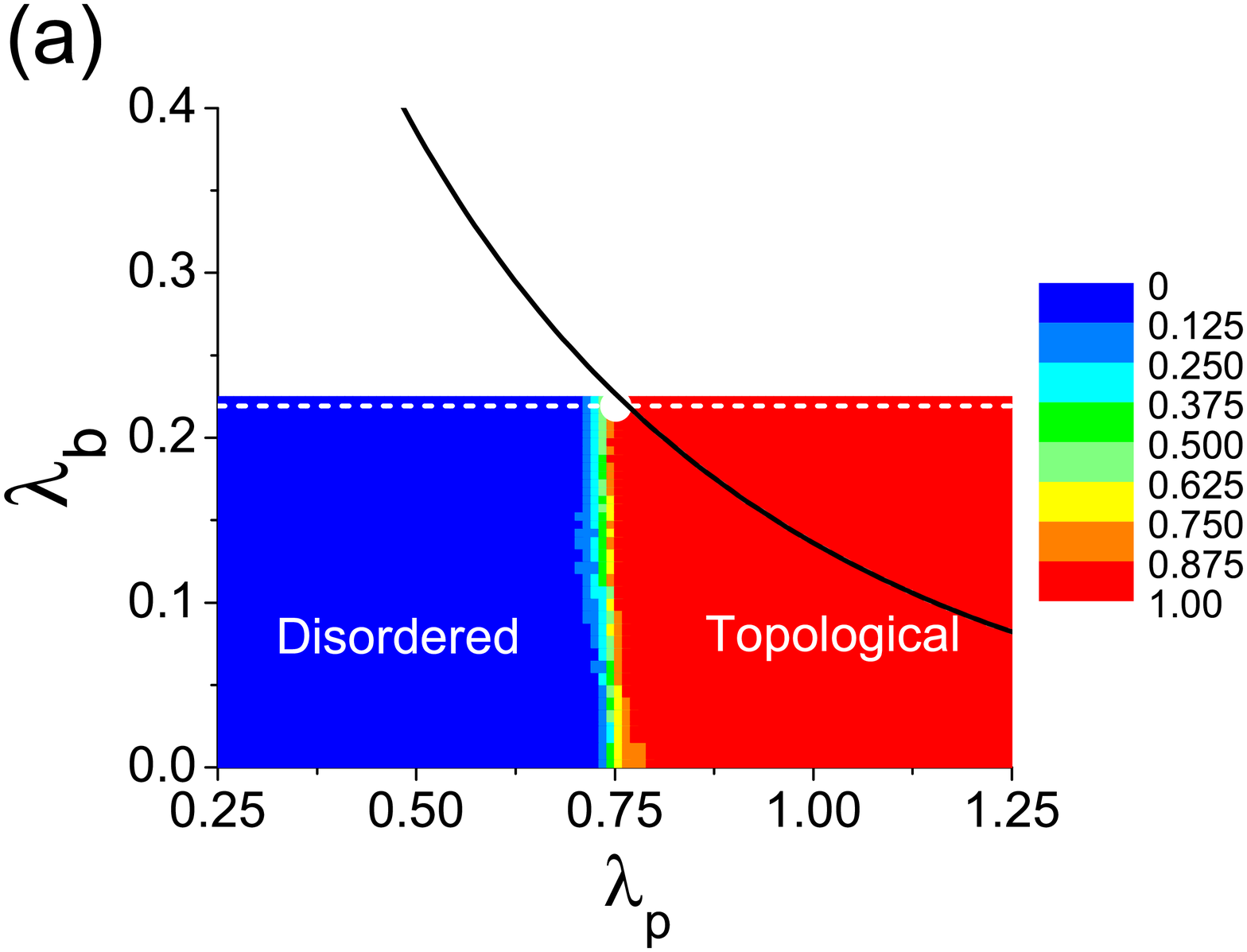}
\includegraphics[scale=0.32]{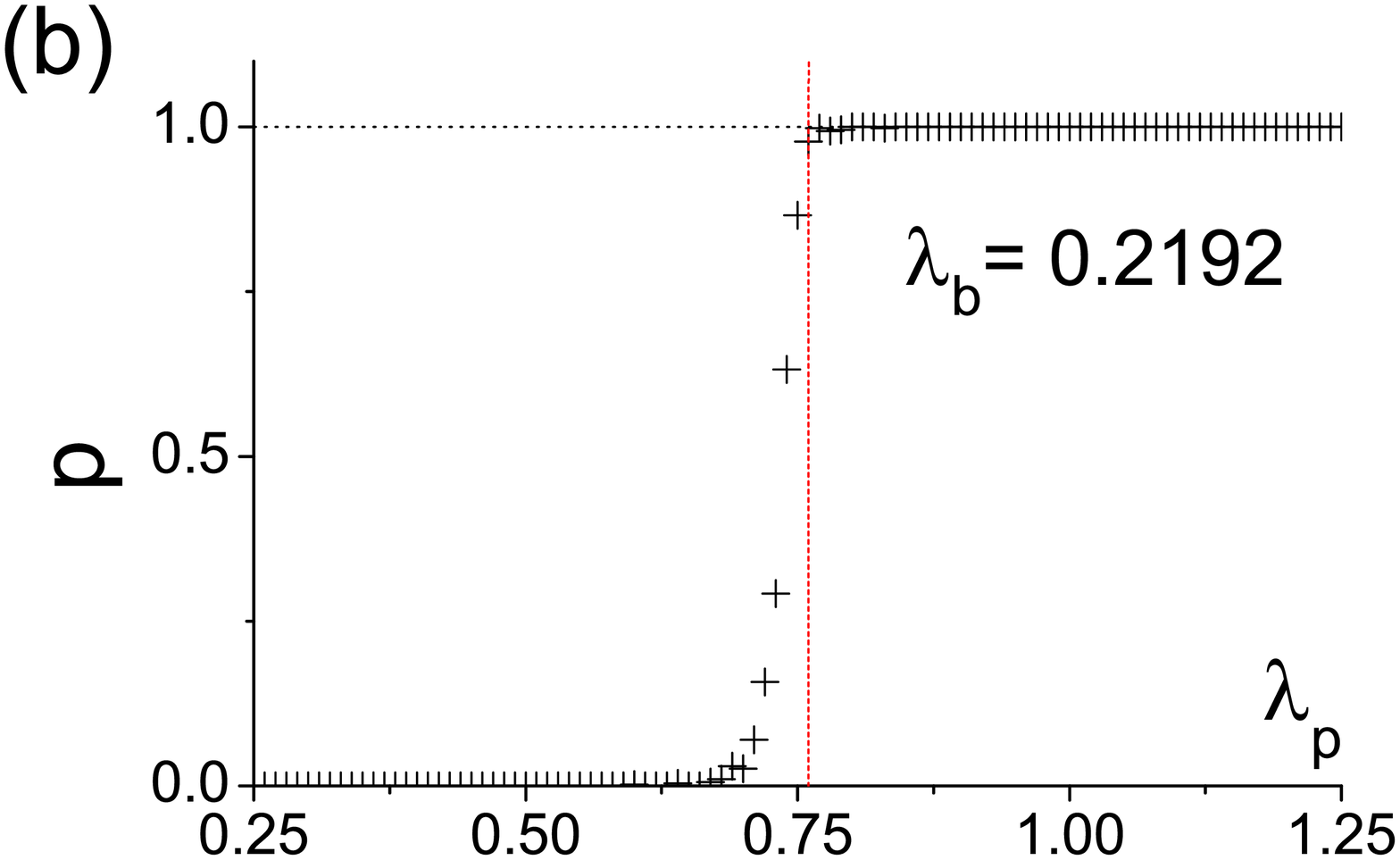}
\caption{(a): For each $\lambda_b \le 0.22$, we implement supervised training with quantum loop topography inputs on the neural network with $\lambda_p=1.25$ for the topological phase ($y=1$) and $\lambda_p=0.25$ for the trivial phase ($y=0$). Then the resulting neural network is applied to the phase space in between and builds up the phase diagram slice by slice. The color scales indicates the fraction $p(\lambda_p, \lambda_b)$ of `topological' output, from red for the most likely to blue for the least likely. The solid black line labels self-dual conditions. (b): one slice at $\lambda_b=0.2192$ corresponding to the white dashed line in (a). The vertical red dashed line in (b) and white dot in (a) label $\lambda_p=0.7553$, the critical value pinpointed using specific heat collapse method, see Fig. \ref{fig:benchmark}. } \label{fig:pd1ds}
\end{figure}

To improve the critical region, we consider a slice-by-slice construction of the phase diagram where a series of neural networks are trained. The resulting phase diagram in Fig. \ref{fig:pd1ds}(a) is obtained with 34 different neural networks. We train one neural network with quantum loop topography inputs from $\lambda_p=1.25$ for the topological phase ($y=1$) and $\lambda_p=0.25$ for the disordered phase ($y=0$), at a fixed value of $\lambda_b\le 0.22$ below the magnetic ordering threshold. Then the phase space interpolating between $0.25\le\lambda_p\le1.25$ is scanned by that particular neural network. To analyze the extent of accuracy we achieve, 2000 QLT test inputs are assessed at each $(\lambda_p, \lambda_b)$ for the fraction $p$ of `topological' response. Indeed, the results that $p\rightarrow 0$ deep in the trivial phase and $p\rightarrow 1$ deep in the topological phase, indicating even a single measurement with as few as five Monte Carlo samples can reliably provide a trustworthy detection. Notice that not only is the critical region relatively sharp, but the phase boundary between the disordered phase and the topological phase now reproduces the slightly negative slope seen in Ref.~\cite{Stamp2010}. Here we only focused on refining the phase boundary between the disordered phase and the topological phase relying on duality to give us the other side of the phase boundary. However, to apply the QLT to unknown phase diagrams, one may consider multiple-neuron output layer to differentiate all phases under equal footing.

We now compare the neural network based location of phase boundary to the critical point on the line $\lambda_b/\lambda_p=0.29$ obtained using specific heat collapse: $(\lambda_p,\lambda_b)=(0.7553\pm0.006,0.2192)$ by considering a cut of the two-dimensional phase diagram along
$\lambda_b = 0.2192$ shown in Fig.~\ref{fig:pd1ds}(b) [see the white dashed line in Fig.~\ref{fig:pd1ds}(a)].
The neural network shows a statistical ratio of `topological' response $p<0.005$ for $\lambda_p<0.7$, deep in the trivial phase, and $p>0.995$ for $\lambda_p>0.8$, deep in the topological phase putting a bound on the critical point $0.7<\lambda_p^c<0.8$. This is clearly consistent with the specific heat collapse guided determination of the critical point, albeit with lower precision. However, we note that the total computational time for the cut shown in Fig.~\ref{fig:pd1ds}(b) was mere 18 minutes of CPU time. Hence, this comparison establishes how the QLT based machine learning can facilitate and speed-up the evaluation of phase diagram dramatically by quickly zooming into the narrow region containing the phase boundary.

\begin{figure}
\includegraphics[scale=0.4]{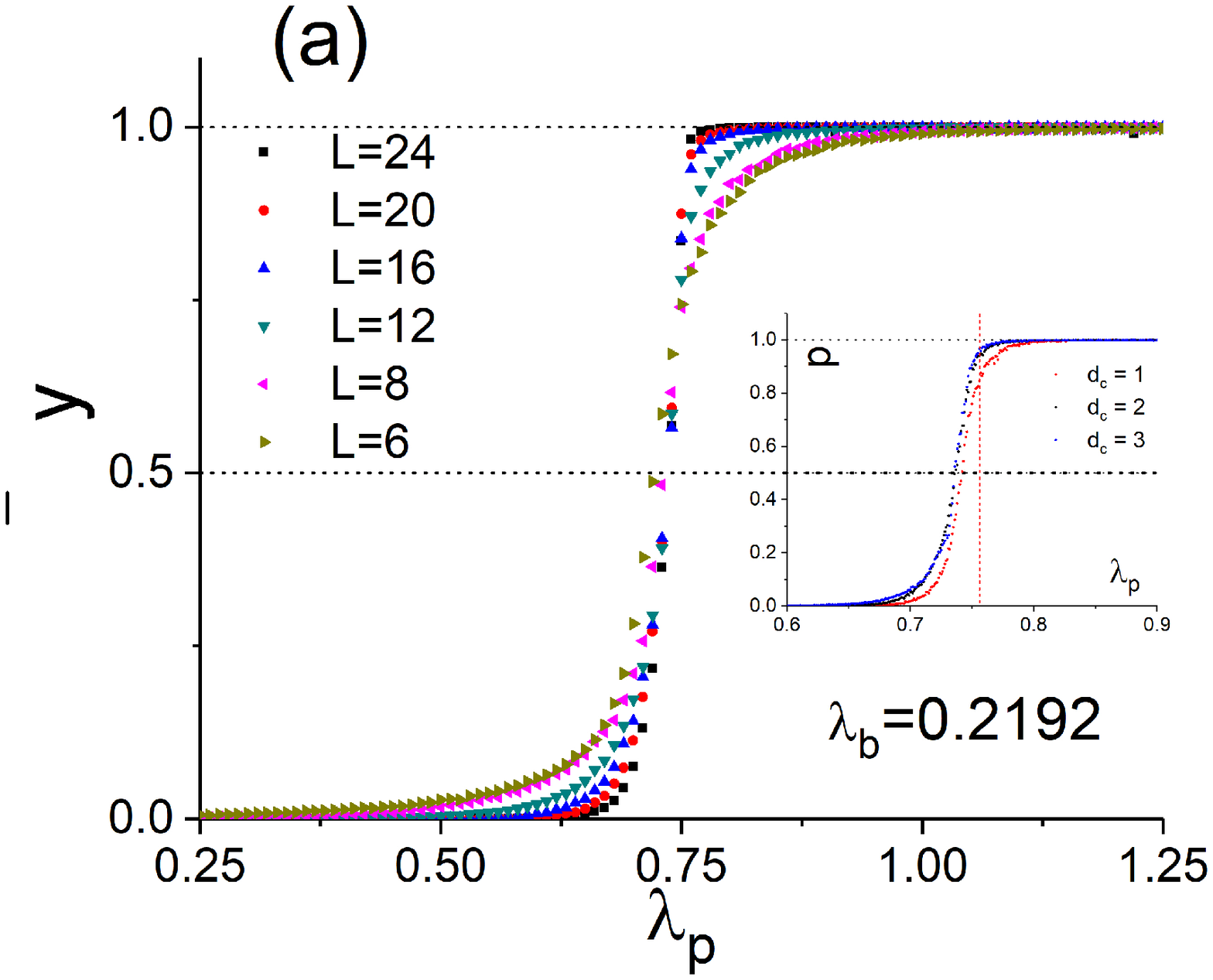}
\includegraphics[scale=0.3]{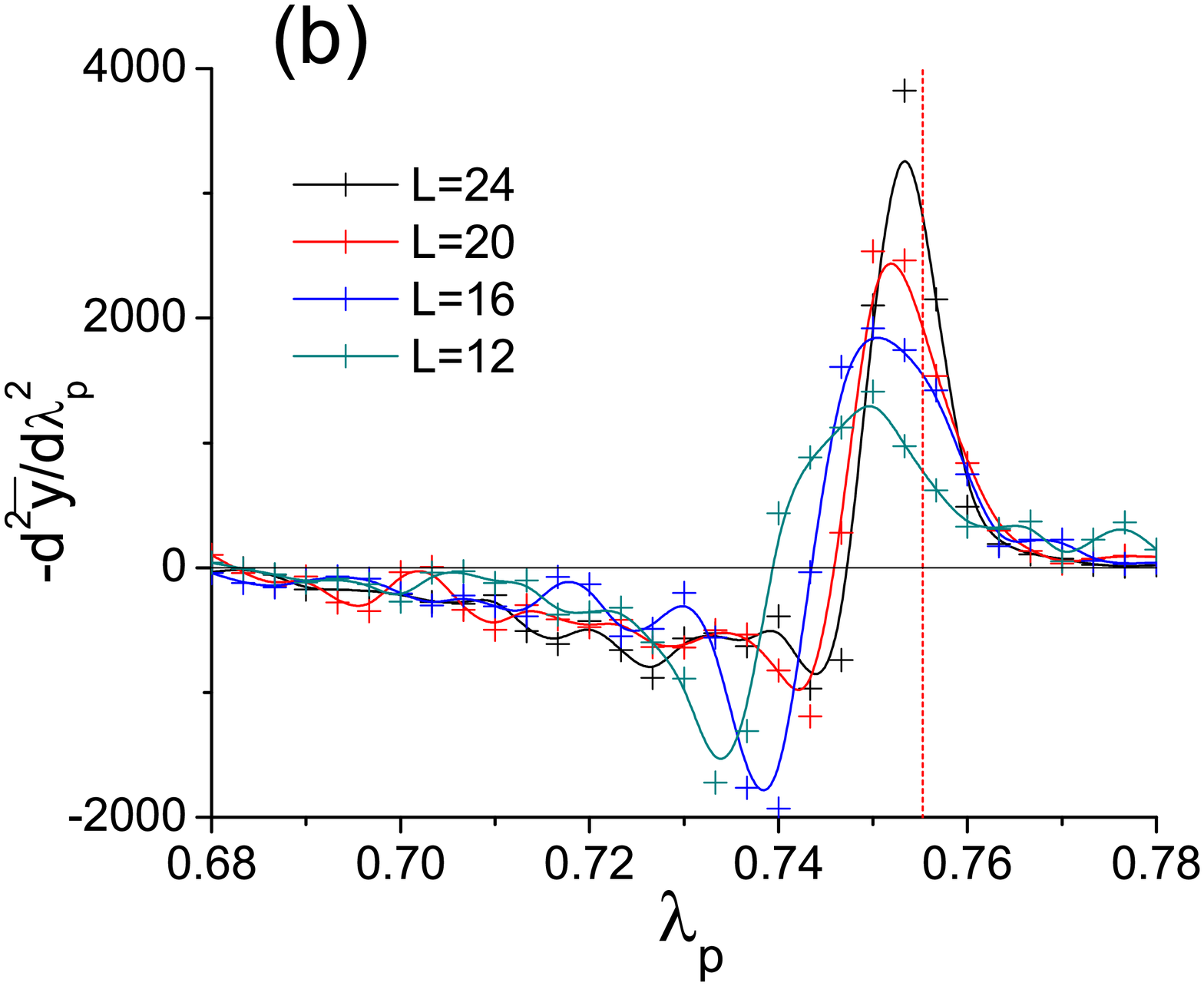}
\caption{(a) The neural output $\bar y(\lambda_p)$ over the range $0.25\le \lambda_p\le1.25$. (b) The second derivative $-d^2\bar y/d\lambda_p^2$ in the critical region $0.68\le \lambda_p\le0.78$ from higher-resolution data. $\bar y$ is averaged over 20000 Monte Carlo samples at $\lambda_b=0.2192$ and different system sizes $L$. Inset: The ratio of topological response `p' for different QLT cut-off $d_c=1,2,3$ at $L=12$. The red dashed lines mark the location of the transition $\lambda_p=0.7553$ obtained using specific heat collapse, see Fig. \ref{fig:benchmark}. }\label{fig:transition}
\end{figure}

Before closing, we now turn to the finite-size effects on the neural network output $\bar y(\lambda_p)$ averaged over 20000 Monte Carlo samples [see Fig. \ref{fig:transition}(a)]. Again for fixed $\lambda_b=0.2192$ for comparison with the specific heat collapse, the training sets are taken from  $\lambda_p=0.25, 0.60$ for the trivial phase and $\lambda_p=0.90, 1.25$ for the topological phase, respectively. The resulting neural network then tests the Monte Carlo samples generated between $0.25\le\lambda_p\le1.25$. We find the outputs to be reliably definitive ($y\rightarrow 0$ in the trivial phase and $y\rightarrow 1$ in the topological phase) away from the critical region even for the smallest system size $L=6$. In addition, no obvious hysteresis is observed. These behaviors are in sharp contrast with the more global probes based upon large Wilson loops \cite{Jongeward1980}. The fact that even rather small systems can be used to detect the topological phase away from the critical region is clearly advantageous. Noticeably, the transition gets sharper with increasing system size in Fig.~\ref{fig:transition}(a). To capture this trend in a more revealing manner we looked into the curvature of $\bar{y}$ as a function of $\lambda_p$. Specifically, we fit the higher-resolution data in the critical region to an analytic curve and take derivatives \footnote{We do not find a discontinuity in the neural output which offered a sharp location of the critical point in Ref.~\cite{qlt2016}. This is possibly due to the fact that the topologically trivial phases are gapless in the present case.}. The data points of Fig.~\ref{fig:transition}(b) are the results and the solid curves are guide to the eyes. This curvature plot clearly highlights how the peak in curvature is sharpening and moving towards the reference point obtained from the specific heat collapse, demonstrating that our machine learning based approach can do a respectable job near criticality, upon increasing system size. We also find a fast convergence in the QLT cut-off $d_c$, see Fig. \ref{fig:transition} inset, and $d_c=2$ is indeed adequate for our exemplary purposes.

\section{Summary and discussions}\label{sec:summary}

We have proposed a quantum-loop-topography-based supervised machine learning strategy for detecting strongly-correlated topological phases, using quasi-particle statistics as defining features. The quantum loop topography (QLT) we developed here is a feature selection process \cite{Guyon2003} that effectively picks out Wilson loop operators on a semi-local scale -- the key feature of a topological phase that is nevertheless minimally extended. Relying on the mutual statistics between the vison and spinon excitations, QLT enabled us to successfully train and test a simple (shallow, fully-connected feed forward) neural network to recognize a $\mathbb Z_2$ quantum spin liquid,  map out its parameter region, and locate the topological quantum phase transitions of a microscopic interacting Hamiltonian.

Compared to existing approaches of studying the phase diagram of $\mathbb Z_2$ quantum spin liquids, the QLT-based machine learning we have implemented here has two clear advantages: (1) significant savings on computational resources as we have summarized in the last section, and (2) targeting small-scale Wilson loops. Indeed, the combination of our QLT and machine learning makes effective use of small-scale Wilson loops reflecting quasi-particle statistics. Traditionally, diagnosis of topological phases and de-confined gauge theories \cite{Sondhi2011njp} has largely relied on long-range entanglement properties in order to avoid intractable short-distance fluctuations. However, we have demonstrated that one can still manage to analyse the information within semi-local operators with neural networks, which excel at being trained to recognize key features even in noisy data. Further, since QLT is semi-local, the approach does not depend on boundary conditions, and even small systems can yield reliable results on labelling the phase. Moreover, this selective nature of QLT preempts memory required for storing and processing data becoming a bottle-neck, which can be an issue with machine learning based approaches.

Going beyond the $\mathbb Z_2$ quantum spin liquids, the model of QLT featuring quasi-particle statistics that we have introduced here can be generalized to other non-trivial topological phases, including ones with non-Abelian excitations. Our development will be particularly beneficial for the study of higher-dimensional systems, such as the three-dimensional $\mathbb{Z}_2$ topological phases \cite{Chamon2005} and Fracton topological order \cite{fracton2016,fracton2017}. The fact that these phases also feature multiple types of quasi-particles with nontrivial statistics and fusion makes them an exciting future direction to study with our algorithm. Finally, we would like to point out that the QLT constructed here and the QLT we constructed in Ref.~\cite{qlt2016} are guided by distinct defining properties of the phases of interest: Hall conductivity for chiral topological phases and mutual statistics for the $\mathbb Z_2$ quantum spin liquid. Nevertheless, both QLT layers enabled successful evaluation of the phase diagrams using a shallow network with orders of magnitude speed up. From the broader perspective of machine learning, the idea of a QLT layer using particular operators to select relevant features as inputs to the neural network is an example of a feature selection layer \cite{Guyon2003}. Our success in Ref.~\cite{qlt2016} and again in this paper advocates QLT -- and the idea of selecting defining features and relevant operators via responses, statistics, or beyond -- as a general strategy for machine learning applied to quantum condensed matter problems. We also note important progress on representing quantum states using neural networks \cite{Carleo2016,Deng2016,Deng2017}.

{\bf Acknowledgements:} We thank Fiona Burnell, Anders Sandvik, Simon Trebst and Paul Ginsparg for useful discussions. We also thank Institute for Theoretical Physics, University of Cologne for hospitality during the Quantum Machine Learning workshop, where some of these ideas were consolidated and the paper was finalized.  YZ acknowledge support through the Bethe Postdoctoral Fellowship and E-AK acknowledges Simons Fellow in Theoretical Physics Award $\#$392182 and DOE support under Award de-sc0010313. Simulations were performed in part on resources provided by SHARCNET through Compute Canada. RGM acknowledges support from NSERC, the Canada Research Chair program, and the Perimeter Institute for Theoretical Physics. Research at Perimeter is supported through Industry Canada and by the Province of Ontario through the Ministry of Research $\&$ Innovation.

\bibliographystyle{apsrev4-1}
\bibliography{refs}

\newpage

\section{Appendix: fluctuation suppression via sample binning}

\begin{figure}
\includegraphics[scale=0.32]{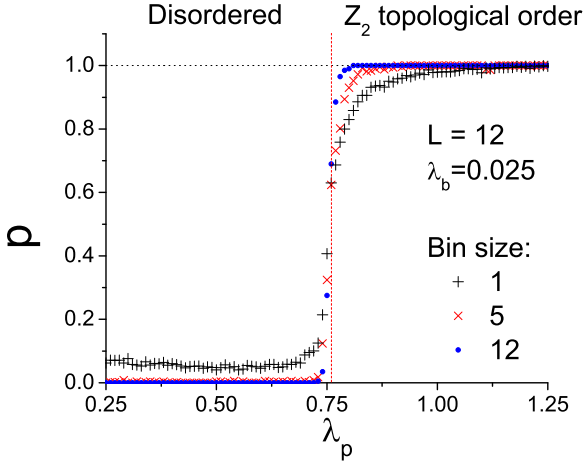}
\caption{The ratio of `topological' response as a function of
$\lambda_p$ at fixed $\lambda_b=0.025$. The neural network is
prepared with training set using $\lambda_p=1.25$ for the
topological phase ($y=1$) and $\lambda_p=0.25$ for the trivial
(disordered) phase ($y=0$), and then applied to the phase space in
between. The decision is based upon the criteria that average neural
output $\bar y>0.5$ over a number of uncorrelated samples. The
vertical red dashed line at $\lambda_p=0.76$ is the approximate
location of the transition.}  \label{fig:pd1d}
\end{figure}

Since our quantum loop topography is built upon individual Monte
Carlo samples and forgoes the Markov chain, the input into and thus
the output $y$ from the artificial neural network naturally have
fluctuations irrespective of the implementation of the supervised
machine learning, and may sometimes influence the decision's
accuracy. For example, using the simple criteria $y>0.5$ to
determine whether a single output is `topological', one can achieve
accuracy of $>99.5\%$ deep in the topological phase
($\lambda_p>0.8$) as well as the trivial phase ($\lambda_p<0.y$) at
relatively large $\lambda_b\ge 0.1$; however, at small $\lambda_b\le
0.05$, such accuracy decreases to $\sim 96\%$, see Fig.
\ref{fig:pd1d}.

To suppress the impacts of the fluctuation, we instead judge
`topological' or `trivial' based upon whether $\bar y>0.5$, where
the neural output is average over a small bin that consists of
multiple Monte Carlo samples. Indeed, with bins as small as 5
samples, accuracy away from the critical region $>99.5\%$ is
obtained once again even at small $\lambda_b$, see Fig.
\ref{fig:pd1d}.

Note that we have kept intact the algorithm of supervised machine
learning and the architecture of the neural network and quantum loop
topography. Instead, our binning only kicks in after we have
obtained our original neural network's outputs, thus retaining full
legitimacy of the discussion and analysis on the supervised machine
learning and quantum loop topography.

\end{document}